\documentclass[11pt]{article}
\newif\iffigs\figstrue

\usepackage{amssymb}

\DeclareMathAlphabet{\mathpzc}{OT1}{pzc}{m}{it}

 \csname
@addtoreset\endcsname{equation}{section}

\def\gz0{\gamma^{0}}

\def\be{\begin{equation}}
\def\ee{\end{equation}}
\def\bea{\begin{eqnarray}}
\def\eea{\end{eqnarray}}
\def\ba{\begin{array}}
\def\ea{\end{array}}
\def\bec{\begin{center}}
\def\ec{\end{center}}
\def\ba{\begin{align}}
\def\ena{\end{align}}

\def\12{\frac{1}{2}}

\usepackage{slashed}

\usepackage{float}

\usepackage{booktabs}

\usepackage[mathscr]{euscript}

\usepackage{color}
\usepackage{graphicx}
\usepackage{epsf}
\usepackage{graphicx,epsfig}
\usepackage{amsmath}

\usepackage{multirow}

\usepackage{amsmath, amsthm, amssymb}

\usepackage{epsfig}
\usepackage{cite}
\usepackage{color,colordvi}

\newcounter{hran}

\makeatletter
\renewcommand\section{\@startsection {section}{1}{\z@}%
                               {-3.5ex \@plus -1ex \@minus -.2ex}%
                               {2.3ex \@plus.2ex}%
                               {\normalfont\large\bfseries}}
\makeatother

\vspace{0.5cm}

\renewcommand{\bea}{\begin{eqnarray}}
\renewcommand{\eea}{\end{eqnarray}}
\newcommand{\bi}{\begin{itemize}}
\newcommand{\ei}{\end{itemize}}

\def\beq{\begin{equation}}
\newcommand{\eeq}[1]{\label{#1}\end{equation}}

\begin{document}
\thispagestyle{empty}
\begin{flushright}
CERN-PH-TH-2015-174\\
CPHT-RR021.0615\\
{\today}
\end{flushright}

\vspace{10pt}

\begin{center}


{\Large\sc Properties of Nilpotent Supergravity}\\


\vspace{25pt}
\begin{center}
{\sc E.~Dudas${}^{\; a,b}$, S.~Ferrara${}^{\; c,d,e}$, A.~Kehagias${}^{\; c,f}$ \ and \ A.~Sagnotti${}^{\; c,g}$}\\[15pt]

{${}^a$\sl\small Centre de Physique Th\'eorique,
\'Ecole Polyt\'echnique\\
F-91128 Palaiseau \ FRANCE\\[4pt]
${}^b$\sl\small Deutsches Elektronen-Synchrotron DESY, 22607 Hamburg \ GERMANY\\[4pt]
}

{${}^c$\sl\small Th-Ph Department, CERN\\
CH - 1211 Geneva 23 \ SWITZERLAND \\[4pt]
${}^d$ INFN - Laboratori Nazionali di Frascati \\
Via Enrico Fermi 40, 00044 Frascati \ ITALY\\[4pt]
${}^e$ Department of Physics and Astronomy,
University of California \\ Los Angeles, CA 90095-1547 \ USA\\[4pt]
}

{${}^f$\sl\small
Physics Division, National Technical University of Athens\\ 15780 Zografou, Athens \ GREECE \\[4pt]
}

{${}^g$\sl\small
Scuola Normale Superiore and INFN\\
Piazza dei Cavalieri \ 7\\ 56126 Pisa \ ITALY \\[12pt]}

{\small \sl
dudas@cpht.polytechnique.fr\\[3pt] sergio.ferrara@cern.ch\\[3pt] kehagias@central.ntua.gr \\[3pt] sagnotti@sns.it}
\end{center}
\vspace{24pt}

\vspace{5pt} {\sc\large Abstract}
\end{center}

\noindent We construct supergravity models where the goldstino multiplet has a gravitational origin, being dual to the chiral curvature superfield. Supersymmetry is nonlinearly realized due to a nilpotent constraint, while the goldstino arises from $\gamma$--traces of the gauge--invariant gravitino field strength. After duality transformations one recovers, as expected, the standard Volkov--Akulov Lagrangian coupled to Supergravity, but the gravitational origin of the goldstino multiplet restricts the available types of matter couplings.
We also construct explicitly some inflationary models of this type, which contain both the inflaton and the nilpotent superfield.

\vskip 40pt
\begin{center}
{\sl Dedicated to the memory of Raymond Stora}
\end{center}

\vfill
\baselineskip=20pt
\noindent

\setcounter{page}{1}

\pagebreak

\newpage

\section{\sc \bf Introduction}

In this note we construct the minimal Supergravity \cite{SUGRA} model with a nilpotent gravitational multiplet, where Supersymmetry is non--linearly realized. In the \emph{old--minimal} formalism~\cite{oldminimal}, this condition results from the constraint
\beq \left(\frac{\cal R}{S_0} \ - \ \lambda\right)^2\ = \ 0 \ , \eeq{int1}
where ${\cal R}$ is the chiral scalar curvature superfield. As we shall see, $\lambda$ is related to the cosmological constant and does not vanish if the theory admits a Minkowski vacuum. The goldstino originates from the gravitino curvature, and at the linearized level in a flat background
\beq
G \ = \ - \ \frac{3}{2\,\lambda} \left( \gamma^{\mu\nu}\, \partial_\mu\, \psi_\nu \ - \ \frac{\lambda}{2} \ \gamma^\mu\, \psi_\mu \right) \ ,
\eeq{int12}
while not surprisingly the dual standard Supergravity is coupled to the Volkov--Akulov theory~\cite{VA}. This model describes the super--Higgs effect~\cite{superhiggs,superhiggs_2,superhiggs_3}, and goldstino modes can be conveniently described via constrained superfields \cite{constrained_superfields_1,constrained_superfields_2,constrained_superfields_3,dumitru}. The universal low--energy couplings to gravity of a massive gravitino depend on two parameters, $W_0$ and $\lambda$, which can be related to a cosmological constant and a mass term. The peculiar feature of the systems resulting from eq.~\eqref{int1} is that \emph{the SuperHiggs effect occurs without an independent field describing the goldstino}. This phenomenon reflects the higher--derivative nature of the gravitino equation, which describes four degrees of freedom consistently with a Stueckelberg realization of the gauge symmetry. It goes on the par with similar phenomena that present themselves in $R+R^2$ theories \cite{grisaru}, where a higher--derivative gravitino equation describes a massless spin-3/2 and two massive spin--1/2 modes.

We then investigate the consistency conditions for coupling the nilpotent Supergravity to chiral multiplets $Q_i$ with linearly realized Supersymmetry, and show that the natural generalization of the nilpotency condition \eqref{int1} is
\beq
\left(\frac{{\cal R}}{S_0} \ - \ f (Q_i)\right)^2\ = \ 0 \ .
\eeq{int2}
The chiral function $f(Q_i)$ is the counterpart of $\lambda$ for these systems: it must acquire a non--zero \emph{v.e.v.} in vacua with broken Supersymmetry and a vanishing cosmological constant, and enters the superpotential $W$ as $f(Q_i) \, X$, where $X$ is the nilpotent goldstino multiplet. We construct explicit examples of positive--definite potentials that admit vacua with a vanishing cosmological constant and where the breaking of Supersymmetry occurs purely in the sgoldstino direction.

The emphasis on a purely gravitational form was partly motivated by nonlinear Supergravity models of inflation, whose avatar, the higher curvature $R+R^2$ Starobinsky model \cite{Starobinsky}, appears currently favored by PLANCK constraints \cite{PLANCK}. Also for this reason, its complete Supergravity embedding \cite{rr2} has been widely discussed in the recent literature, in both the \emph{old minimal} \cite{oldminimal} and \emph{new minimal} \cite{newminimal} formulations.

Constrained chiral superfields were introduced, in the higher--curvature old minimal Volkov--Akulov--Starobinsky Supergravity, in \cite{adfs}. In a different context, still related to inflation, they were previously considered in \cite{Alvarez-Gaume}, and more recently the construction was extended further in \cite{sugra_inflation,sugra_inflation_1,sugra_inflation_2}. In \cite{adfs}, the chiral curvature multiplet ${\cal R}$ was subject to the constraint ${\cal R}^2=0$, which translated into the corresponding constraint $X^2=0$ for the goldstino multiplet, one of the two chiral multiplets present in the two--derivative dual Supergravity.

Non--linear realizations of Supersymmetry play an important role in String Theory \cite{stringtheory}, in orientifold vacua \cite{orientifolds} where a  high--scale breaking is induced by mutually non--supersymmetric collections of branes and orientifolds. While this setting, usually termed ``brane Supersymmetry breaking'' \cite{bsb}, brings non--linear realizations of Supersymmetry to the forefront at a potentially deeper level \cite{dm}, its four--dimensional counterparts in the low--energy Supergravity fall nicely into the class of models reviewed here. They bear a close relationship to the KKLT uplift \cite{kklt}, as discussed in \cite{dks} and, more recently, in \cite{kallosh}.

This paper is organized as follows. In Section \ref{sec:nls} we introduce the minimal nilpotent Supergravity and discuss in detail the corresponding higher--derivative gauge--invariant description of a massive gravitino. In Section \ref{sec:dcf} we show that this formulation is dual to the Volkov--Akulov model coupled to standard Supergravity, which embodies the conventional gauge--invariant description of a massive gravitino \cite{superhiggs,superhiggs_2,superhiggs_3}. In Section \ref{sec:matter} we discuss the matter couplings to ordinary scalar multiplets that are allowed in the gravitational formulation and show that, in the dual two--derivative formulation, they give rise to \emph{restricted couplings} to the nilpotent Volkov--Akulov multiplet.  We also present some explicit examples of no--scale models of this type \cite{noscale,noscale_2}, which combine semi--definite potentials and broken Supersymmetry.
Finally, we describe a class of inflationary models with gravitational duals, which combine the minimal non--linear Supergravity with an inflaton multiplet whose the K\"ahler potential possesses a shift symmetry.

\section{\sc  \bf Minimal nilpotent Supergravity}\label{sec:nls}

The minimal Supergravity action is \cite{SUGRA}
\begin{equation}
{\cal L} \ = \ \left[- S_0 \, {\overline S}_0  \right]_D \ + \ \left[W_0 S_0^3 \right]_F \ , \label{staro1}
\end{equation}
where the subscripts identify $D$ and $F$ superspace densities and $S_0$ is the chiral compensator field. In components and in a mostly positive signature, the complete action of eq.~\eqref{staro1} becomes
\begin{eqnarray}
{\cal S} &=& \int d^4 x \ e\, \left\{ \ - \ \frac{1}{2}\
R \ - \  \frac{i}{2}\ {\bar \psi}_\mu \ \gamma^{\mu\nu\rho}\ {D_\nu \psi_\rho} \ + \ \frac{1}{3} \ A_\mu \, A^\mu  \right. \nonumber \\ &-& \frac{1}{3}\
u \, {\bar u} \ + \ \left. W_0 \left( u \ +\  \frac{i}{2} \ {\bar \psi}_{\mu\, R} \ \gamma^{\mu\nu} \ \psi_{\nu\, R} \ + \  {\rm h.c.} \right)  \right\} \ , \label{nls1}
\end{eqnarray}
where $u$ and $A_m$ are the auxiliary fields of the old minimal \cite{oldminimal} Supergravity. This Lagrangian is invariant under the off--shell supersymmetry transformations given in \cite{superhiggs_3}.

In order to enforce a nonlinear realization of Supersymmetry, in analogy with \cite{adfs} we impose the chiral constraint
\be
\left(\frac{\cal R}{S_0} \ -\  \lambda\right)^2 \ = \ 0 \  , \label{nls2}
\ee
where $ {\cal R}$ is the chiral scalar curvature superfield, defined via the curved chiral projector $\Sigma$ as~\cite{SUGRA}
\be
{\cal R} \ = \ \frac{\Sigma ({\overline S}_0)}{S_0} \ ,\label{nls3}
\ee
and $\lambda$ is a constant whose interpretation we are about to describe.

Notice that eq.~\eqref{nls2} is a supersymmetric constraint, and yet the introduction of $\lambda$ results in the spontaneous breaking of Supersymmetry. As a result, one is building a Stueckelberg realization without an independent goldstino field, which makes this setting somewhat unusual. The vacuum energy of the nonlinear supergravity, obtained substituting in eq.~\eqref{nls1} the field independent part of the solution of the constraint~\eqref{nls2}, $\overline{u}=\lambda$, is
\begin{equation}
V \ = \ \frac{\lambda}{3}\, \left( \lambda \ - \ 6\, W_0 \right) \ = \ \frac{1}{3} \left(\lambda \ - \ 3\, W_0 \right)^2 \ - \ 3\, W_0^2 \ . \label{potential}
\end{equation}
Therefore, any sign is allowed, and in particular the Minkowski case with broken Supersymmetry obtains if $\lambda = 6 W_0$.

Taking into account the nilpotency of the fermionic component, one can see that the cubic constraint for $u$ resulting from eq.~\eqref{nls2} reduces to the quadratic equation
\be
\left({\overline u} \ - \ \lambda \right) \left( \ - \ \widehat{R}
\,+\, \frac{2}{3} \ A_\mu^{\,2} \,-\, 2\, i \ \widehat{{\cal D}^\mu A_\mu} \ + \ \frac{4}{3} \ \lambda \  u \, \right) \ = \ 2\, i \ \widehat{{\cal D}_{\mu}\, \overline{\psi}_{\nu\, L}}\ \gamma^{\mu\nu} \, \gamma^{\rho\sigma} \, \widehat{{\cal D}_\rho \, \psi_{\sigma\,L}}  \ , \label{nls5}
\ee
where ``hats'' denote supercovariant extensions, which are computed at $\overline{u}=\lambda$. The complete solution for $u$,
\begin{equation}
\overline{u} \ - \lambda \ = \ \frac{2\, i \ \widehat{{\cal D}_{\mu}\, \overline{\psi}_{\nu\, L}}\ \gamma^{\mu\nu} \, \gamma^{\rho\sigma} \, \widehat{{\cal D}_\rho \, \psi_{\sigma\,L}}}{\alpha \ + \ \frac{4\, \lambda^2}{3}} \left( 1 \ - \ \frac{8\, i\, \lambda}{3} \ \frac{\widehat{{\cal D}_{\mu}\, \overline{\psi}_{\nu\, R}}\ \gamma^{\mu\nu} \, \gamma^{\rho\sigma} \, \widehat{{\cal D}_\rho \, \psi_{\sigma\,R}}}{\left| \alpha \ + \ \frac{4\, \lambda^2}{3} \right|^2} \right)
\end{equation}
where
\begin{equation}
\alpha \ =  \ - \ \widehat{R} \ + \ \frac{2}{3} \ A_\mu^2 \ - \ 2\, i \ \widehat{{\cal D}^\mu A_\mu} \ ,
\end{equation}
is actually relatively simple, due to the nilpotency of the goldstino bilinears.

One can illustrate the amusing properties of this system rather clearly in a Minkowski vacuum, while confining the attention to terms that are at most quadratic in the gravitino.
In this case the relevant parts of local Supersymmetry variations in \cite{superhiggs_3}, which do not involve the auxiliary field $A_\mu$, read
\begin{eqnarray}
&& \delta\, \psi_{\mu\, L} \ = \ \partial_\mu \, \epsilon_L \ + \ \frac{\overline{u}}{6} \ \gamma_\mu \, \epsilon_R \ , \nonumber \\
&& \delta\, u \ = \ i\, \overline{\epsilon}_R \ \gamma^{\mu\nu} \ \partial_\mu \, \psi_{\nu\, R} \ - \ \frac{i\, u}{2} \ \overline{\epsilon}_R\, \gamma^\mu \, \psi_{\nu\, L}
\label{nls75}
\end{eqnarray}
where, to lowest order, one can set $u=\lambda$. To quadratic order in the Fermi fields, the solution of the constraint (\ref{nls2}) is then
\begin{equation}
u \ = \ \lambda \ - \ \frac{2\, i}{3}\ \overline{G}_R \ G_R \ , \label{nls6}
\end{equation}
where in this limit the goldstino field is
\begin{equation}
G \ = \ - \ \frac{3}{2\,\lambda} \left( \gamma^{\mu\nu}\, \partial_\mu\, \psi_\nu \ - \ \frac{\lambda}{2} \ \gamma^\mu\, \psi_\mu \right) \ . \label{nsl62}
\end{equation}

As we anticipated, in this setting \emph{the goldstino is not an independent field}. Rather, it results from the application to the gravitino of a differential operator. It shifts properly under a supersymmetry transformation, since
\begin{equation}
\delta\, G \ = \ \frac{\lambda}{2} \ \epsilon \ ,
\end{equation}
while the correction term in eq.~\eqref{nls6} is fully determined by the linearized off--shell supersymmetry transformations \eqref{nls75}. Similar considerations would apply in any background with broken Supersymmetry, with $\lambda \neq 3\, W_0$ and the cosmological constant as given in eq.~\eqref{potential}, where the lowest--order transformation of the corresponding goldstino would read
\begin{equation}
\delta\, G \ = \ \left(\lambda \ - \ 3\, W_0 \right) \, \epsilon \ .
\end{equation}

The quadratic fermionic terms in the Lagrangian add up to an unconventional Stueckelberg realization of broken Supersymmetry, encoded by
\begin{equation}
{\cal S} \ = \ \int d^4 x \ \left\{ \ - \ \frac{i}{2}\ {\bar \psi}_\mu \ \gamma^{\mu\nu\rho} \ \partial_\nu \, \psi_\rho \ +\  \frac{i\, \lambda}{12} \ {\bar \psi}_{\mu} \ \gamma^{\mu\nu} \ \psi_{\nu} \ + \  \frac{i\, \lambda}{9} \ \overline{G} \ G  \right\} \ , \label{nls21}
\end{equation}
where the last term introduces in the Rarita--Schwinger equation some contributions involving up to two derivatives. The higher--derivative terms are instrumental to grant the invariance under the gauge transformations~\eqref{nls75} of the Lagrangian \eqref{nls21}, whose variation produces terms of type $\overline{\psi} \, \partial \epsilon$ and $\overline{\psi} \, \epsilon$, which must cancel separately. The latter originate from the second and third terms, while the former originate from all of them. As we shall see in the next section, a duality transformation turns this unconventional gauge invariant description of a massive gravitino into a more conventional gauge invariant coupling of the Volkov--Akulov model to Supergravity. It is known that the latter can be formulated as a standard coupling to Supergravity \cite{adfs,Alvarez-Gaume,sugra_inflation,sugra_inflation_1,sugra_inflation_2} of a nilpotent chiral superfield \cite{constrained_superfields_1, constrained_superfields_2,constrained_superfields_3}.

It is instructive to take a closer look at the modified Rarita--Schwinger equation following from the action \eqref{nls21}, which reads
\begin{equation}
\gamma^{\mu\nu\rho}\, \partial_\nu\, \psi_\rho \ - \ \frac{\lambda}{6}\ \gamma^{\mu\nu}\, \psi_\nu \ 1 \ \frac{1}{3} \left(\gamma^{\mu\nu}\, \partial_\nu\ - \frac{\lambda}{2} \ \gamma^\mu\right) G \ = \ 0 \ ,
\end{equation}
The $\gamma$--trace and the divergence of this equation lead to the same gauge--invariant constraint
\begin{equation}
\gamma^{\mu\nu}\, \partial_\mu\, \psi_\nu \ - \ \gamma^\mu\, \partial_\mu \ G \ = 0 \ ,
\end{equation}
where the gaugino $G$ is defined in eq.~\eqref{nsl62}. Gauging away $G$ one thus recovers the two standard constraints
\begin{equation}
\gamma^{\mu\nu}\, \partial_\mu\, \psi_\nu \ = \ 0 \ , \qquad \gamma^\mu\, \psi_\mu \ = \ 0 \ ,
\end{equation}
together with the conventional field equation for a gravitino of mass $W_0=\frac{\lambda}{6}$.

As we have stressed in the Introduction, there is an interesting analogy between this type of construction and those occurring in $R+R^2$ theories \cite{grisaru}. Actually, the analogies go even further, since similar constructs enter also the higher--spin equations of \cite{geom_high}, where higher--derivative compensators built from the original Fronsdal field \cite{fronsdal} result in geometric expressions involving higher--spin curvatures.

\section{\sc  \bf Dual standard Supergravity formulation }\label{sec:dcf}

The action (\ref{nls1}), supplemented with the constraint (\ref{nls2}), can be recast in a two--derivative dual form proceeding along the lines of \cite{rr2}.
To this end, one starts from a Hubbard--Stratonovich transformation, which allows to recast the higher--derivative Lagrangian (\ref{staro1}), (\ref{nls2}) in the form
\be
{\cal L} \ = \ \bigg[- \ S_0 \, {\overline S}_0  \bigg]_D \ + \
\left[  \left\{ X \left(\lambda \ - \ \frac{\cal R}{S_0} \right) \ - \ \frac{1}{4\Lambda_1} \ X^2 \ + \ W_0 \right\} S_0^3 \right]_F \ , \label{aug1}
\ee
and then  into
\begin{equation}
{\cal L} \ = \ \bigg[\ - \ \left(1 \ +\  X \ + \ {\bar X}\right) \ S_0 \, {\overline S}_0  \bigg]_D \ + \
\left[  \left\{ \lambda \, X \ - \ \frac{1}{4\Lambda_1} \ X^2 \ + \ W_0 \right\} S_0^3 \right]_F  \  \label{aug2}
\end{equation}
via the identity  \cite{rr2}
\begin{equation}
\bigg[ f(\Lambda)\ {\cal R}\ S_0^2 \bigg]_F \ + \ {\rm h.c.} \ = \ \bigg[ \big(f(\Lambda) \ + \ {\overline f} ({\overline \Lambda})\big)
\, S_0 \, {\overline S}_0  \bigg]_D  \ + \ {\rm tot. \ deriv.} \ , \label{aug3}
\end{equation}
which holds for any chiral superfield $\Lambda$ and for any function $f$.

That the Lagrange multiplier $\Lambda_1$ does not introduce additional
degrees of freedom is indicated by the linearized analysis of the equations of motion, but its role becomes more transparent in the dual formulation. In this case its field equation imposes a nilpotency constraint on $X$, so that one is finally led to a standard $N=1$ Supergravity with K\"ahler potential $K$ and superpotential $W$ given by
\begin{equation}
K \ =\  - \, 3 \, \ln \left(1 + X \, + \, {\overline X} \right) \ , \qquad W \, \ = \ W_0
\ + \  \lambda \, X  \ , \label{aug3}
\end{equation}
where $X^2=0$.
 After a Taylor expansion and a K\"ahler transformation, one is finally led to
\be
K \ = \ 3 \, |X|^2 \ , \qquad W \ = \ W_0 \ + \ (\lambda\ - \ 3\, W_0) X  \ , \label{chiral11}
\ee
with $X$ satisfying again the constraint $X^2=0$. This defines a dual Volkov-Akulov supergravity action, with a supersymmetry breaking parameter
\begin{equation}
f \, = \, \lambda\, - \, 3\, W_0 \ .
\end{equation}

It is instructive to see these steps in detail reconsidering the quadratic action \eqref{nls21} and introducing an independent goldstino field $\chi$ via a Hubbard--Stratonovich transformation, which turns it into
\begin{eqnarray}
{\cal S} &=& \int d^4 x \ \left\{ \ - \ \frac{i}{2}\ {\bar \psi}_\mu \, \gamma^{\mu\nu\rho}\, \partial_\nu \, \psi_\rho \ +\  \frac{i\, \lambda}{12} \ {\bar \psi}_{\mu} \, \gamma^{\mu\nu} \, \psi_{\nu} \ - \  \frac{2\, i\, \lambda}{3} \ \overline{\chi} \, \chi  \right. \nonumber \\
&-& \left. i\, \sqrt{\frac{2}{3}} \ \overline{\chi} \left( \gamma^{\mu\nu} \, \partial_\mu\, \psi_\nu \ - \ \frac{\lambda}{2} \ \gamma^\mu \, \psi_\mu \right)
\right\} \ . \label{chiral13}
\end{eqnarray}
One can now diagonalize the kinetic terms redefining $\psi_\mu$ according to
\begin{equation}
\psi_{\mu} \ \longrightarrow \ \psi_{\mu} \ + \ \frac{1}{\sqrt{6}} \ \gamma_{\mu} \, \chi \ , \label{chiral14}
\end{equation}
and the end result is a canonical presentation of the superHiggs mechanism \cite{superhiggs,superhiggs_2,superhiggs_3}, described by
\begin{eqnarray}
{\cal S} &=& \int d^4 x \ \left\{ \ - \ \frac{i}{2}\ {\bar \psi}_\mu \, \gamma^{\mu\nu\rho}\, \partial_\nu \, \psi_\rho \ +\  \frac{i\, \lambda}{12} \ {\bar \psi}_{\mu} \, \gamma^{\mu\nu} \, \psi_{\nu} \ - \  \frac{i\, \lambda}{6} \ \overline{\chi} \, \chi  \right. \nonumber \\
&-& \left. \frac{i}{2}\ {\bar \chi} \, \gamma^{\mu}\, \partial_\mu \, \chi \ + \
\frac{i\, \lambda}{2\,\sqrt{6}} \ \overline{\chi} \, \gamma^\mu \, \psi_\mu
\right\} \ . \label{chiral15}
\end{eqnarray}

This is precisely the standard Volkov--Akulov supergravity at the quadratic order in fermions, and as we have seen the duality extends to the full non--linear actions.
Let us stress again that a vacuum with broken Supersymmetry, for which $D_X W \not=0$, is compatible with a vanishing cosmological constant only if $\lambda = 6 \, W_0$, with both $\lambda$ and $W_0$ not vanishing, so that $W_0$ is the gravitino mass.

\section{ \sc \bf Nilpotent Supergravity coupled to matter}\label{sec:matter}

We can now couple the nonlinear Supergravity introduced in the preceding section to standard unconstrained chiral multiplets $Q_i$, with an arbitrary K\"ahler potential $K$ and an arbitrary superpotential $W$. As we shall see shortly, the gravitational origin of these models leaves an interesting imprint in their two--derivative formulation. The Lagrangians of these models are of the form
\begin{equation}
{\cal L} \ = \ \left[\ - \ e^{\,- \,\frac{1}{3} \,K_0 (Q_i, {\bar Q}_{\bar i})} \ S_0 \, {\overline S}_0  \right]_D \ + \ \left[W_0 (Q_i) \, S_0^3 \right]_F \ , \label{m1}
\end{equation}
and are to be supplemented by the nilpotent constraint
\begin{equation}
\left(\frac{{\cal R}}{S_0} \ - \ f (Q_i)\right)^2\ = \ 0 \ , \label{m2}
\end{equation}
where $f (Q_i)$ is a holomorphic function. This expression is a natural generalization of the constraint for the nonlinear Supergravity of the preceding sections in the presence of a number of standard chiral multiplets, if Supersymmetry is still nonlinearly realized only in the gravity multiplet.

Using the same Lagrange multipliers as in Section \ref{sec:dcf}, one can recast eqs.~(\ref{m1}) and (\ref{m2}) in the form
\begin{eqnarray}
{\cal L} &=& \bigg[-  e^{\,- \,\frac{1}{3} \, K_0 (Q_i, {\bar Q}_{\bar i})} \ S_0 \, {\overline S}_0  \bigg]_D \ + \
\left[  \left\{ \Lambda_1 \left(\frac{\cal R}{S_0} - f(Q_i) \right)^2 \, + \, W_0 (Q_i )\right\} S_0^3 \right]_F \label{m3} \\
&=& \bigg[-  e^{\,- \,\frac{1}{3}\, K_0 (Q_i, {\bar Q}_{\bar i})} \ S_0 \, {\overline S}_0  \bigg]_D \ + \ \left[  \left\{  X \left( f(Q_i) - \frac{\cal R}{S_0} \right) -  \frac{1}{4\Lambda_1} X^2 \, + \,
W_0 (Q_i) \right\} S_0^3 \right]_F \ ,\nonumber
\end{eqnarray}
while the identity (\ref{aug3}) turns the Lagrangian \eqref{m1} into
\begin{equation}
{\cal L} = \bigg[- \left( e^{- \frac{1}{3} \,K_0 (Q_i, {\bar Q}_{\bar i})}+ X +  {\bar X}\right) S_0 \, {\overline S}_0  \bigg]_D \ + \
\left[  \left\{ f(Q_i)\, X \, - \frac{1}{4\Lambda_1} \, X^2 \,+\, W_0 (Q_i) \right\} S_0^{\,3} \right]_F  \ . \label{m4}
\end{equation}
The field equation for the unconstrained Lagrange multiplier superfield $\Lambda_1$ now imposes the constraint $X^2=0$, and one is finally led to a standard $N=1$ Supergravity with K\"ahler potential $K$ and superpotential $W$ given by
\begin{eqnarray}
K &=& \ - \, 3 \, \ln \left(e^{- \frac{1}{3} K_0 (Q_i, {\bar Q}_{\bar i})} \ + \ X \ + \ {\overline X} \right) \nonumber \\
 &=& K_0 (Q_i, {\bar Q}_{\bar i}) \ - \ 3\,  e^{ \,\frac{1}{3}\, K_0 (Q_i, {\bar Q}_{\bar i})} (X + {\bar X}) \ + \ 3\, e^{\,\frac{2}{3} \,K_0 (Q_i, {\bar Q}_{\bar i})} \ X \,{\bar X} \ , \nonumber \\
W &=& W_0 (Q_i)
\ + \  f (Q_i) \, X  \ , \label{m5}
\end{eqnarray}
where $X$ is always subject to the quadratic constraint $X^2=0$.

It is now convenient to define the quantity
\begin{equation}
a^{-1} \ = \ 3 \ - \ K_{0,{\bar i}} \ K_0^{{\bar i} j} \ K_{0,j} \ , \label{m6}
\end{equation}
since it is then straightforward to show, introducing the chiral indices $I = (i,X)$ and the corresponding K\"ahler metric $K_{I \bar J}$, that
\begin{equation}
Det \ K_{I \bar J} \ = \ a^{-1} \, e^{\, \frac{2}{3} \, K_0 (Q_i, {\bar Q}_{\bar i})} \ Det \ K_{0, i \bar j}
\ . \label{m7}
\end{equation}
The theory is thus consistent insofar as
\begin{equation}
a \ > \ 0 \quad \longrightarrow \quad  K_{0,{\bar i}} \, K_0^{{\bar i} j} \, K_{0,j} \ < \ 3 \ . \label{m8}
\end{equation}
It is also convenient to introduce the vector
\begin{equation}
V^i \ = \ K_{0, \bar j} \ K_0^{{\bar j} i} \ ,
\end{equation}
since it can be shown that the inverse K\"ahler metric can be cast in the form
\begin{eqnarray}
K^{{\bar I} J} =
\begin{pmatrix}
K_0^{{\bar i} j} \ + \ a\, V^{\bar i}\, V^j & a \, e^{- \frac{1}{3} \, K_0} \, V^{\bar i}  \\
a \ e^{ - \frac{1}{3} \, K_0} \, V^{j} &  a \ e^{- \frac{2}{3} \, K_0}
\end{pmatrix}  \ . \label{m9}
\end{eqnarray}
The corresponding scalar potential is
\begin{eqnarray}
 V &=& e^{K_0} \left\{ K_0^{{\bar i} j} \, \overline{D_i W} \, D_j W \ + \ a\ |V^i D_i W|^2
\ + \ a \ e^{\,- \,\frac{1}{3} \, K_0}  \left( \overline{V^j D_j W} \ D_X W \ + \ {\rm h.c.} \right) \right. \nonumber \\
&+& \left.\ a \ e^{\,- \,\frac{2}{3}\, K_0} \, |D_X W|^2 \ - \ 3\, |W_0|^2 \right\}  \ , \label{m10}
\end{eqnarray}
where we used the nilpotency condition $X^2=0$, and where
\begin{equation}
D_i W \ = \ W_{0,i} \ + \ K_{0,i} W_0 \ , \quad D_X W \ = \ f \ - \ 3 \, e^{\,\frac{1}{3}\, K_0} \ W_0
\ . \label{m11}
\end{equation}
This expression can be cast in the simpler form
\begin{eqnarray}
V &=& e^{K_0} \bigg\{ K_0^{{\bar i} j} \ \bar{W}_{0,\bar{i}} \ W_{0,j} \ + \ a \ e^{-
\frac{2}{3} \,K_0} \left|f \ + \
e^{\frac{1}{3}\, K_0} \,K_0^{{\bar i} j} \,K_{0,\bar i}\  W_{0,j}  \right|^2  \nonumber \\
 &-&  e^{- \frac{1}{3} \, K_0}  ({\bar f}\, W_0 \ + \ f \,{\bar W_0})  \bigg\}  \ . \label{m12}
\end{eqnarray}

As we have stressed, matter multiplets should satisfy the condition $a >0$. One can then see that, in the particular cases
$f=0$ or $W_0=0$, the potential is positive definite. In general, the cosmological constant
can have either sign, but a vanishing cosmological constant where Supersymmetry breaking originates solely from the nilpotent superfield X, as enforced by the conditions $D_i W=0$, $D_X W \not=0$, implies that
\begin{equation}
\langle f \rangle \ = \ \left( \, 3 \ \pm \ \sqrt{\frac{3}{a}}\, \right) \ \langle \ W_0 \ e^{\,\frac{1}{3}\, K_0} \  \rangle  \ . \label{m13}
\end{equation}
In this Minkowski vacuum with broken Supersymmetry, the analog of the Deser-Zumino relation between the scale of Supersymmetry breaking and the gravitino mass is
\begin{equation}
\langle D_X W \rangle \ = \ \pm \ \sqrt{\frac{3}{a}} \ \langle e^{\,\frac{1}{3}\, K_0} \ W_0 \rangle
\ . \label{m14}
\end{equation}
Notice that the chiral function of the matter fields $f(Q_i)$, with a nonzero vacuum expectation value, is crucial in order to attain Supersymmetry breaking with a vanishing cosmological constant. It is the counterpart, in this more general class of models, of the parameter $\lambda$ that we introduced in the preceding sections.

We can now explain the claim concerning the imprints of the gravitational origin of the models. As stressed in \cite{sugra_inflation_2}, the general Lagrangian coupling the nilpotent chiral superfield $X^2=0$ to chiral matter would rest on
\begin{eqnarray}
&& K  \ = \ K_0 (Q_i, {\bar Q}_{\bar i} ) \ + \ \left( K_1 (Q_i, {\bar Q}_{\bar i} ) X \ + \ {\rm h.c.} \right) \ + \ K_2 (Q_i, {\bar Q}_{\bar i}) \, X \, {\bar X} \ , \nonumber \\
&& W \, \ = \ W_0 (Q_i) \ + \  f (Q_i) \, X  \ , \label{m15}
\end{eqnarray}
with $K_1$, $K_2$, $K_3$ and $f$ independent functions. These expressions are clearly more general than eqs.~\eqref{m5}, which arise, as we have seen, from a gravitational constraint of the type (\ref{m2}). The equivalence is possible only when $K_0$, $K_1$ and $K_2$ are related as in (\ref{m5}), so that only this subclass of models possesses a gravitational origin. Conversely, even when the models do not have a gravitational origin, there is always a consistency condition for the action (\ref{m15}), since one can write
\begin{equation}
Det \ K_{I \bar J} \ = \ \left( K_2 \ - \ K_{1,{\bar i}} \ K_0^{{\bar i} j} \ {\bar K}_{1,j} \right) \
Det \ K_{0, i \bar j} \ > \ 0 \longrightarrow \quad  K_{1,{\bar i}} \, K_0^{{\bar i} j} \, {\bar K}_{1,j} \ < \ K_2
\ . \label{m16}
\end{equation}
%
\subsection{ \sc \bf Positive definite potentials}\label{sec:potentials}

Simple examples with zero cosmological constant can be constructed starting from the matter K\"ahler potential
\begin{equation}
K_0 \ = \ - \ 3 \ \ln \left(1 \ - \ |Q|^2\right) \ , \label{noscale1}
\end{equation}
where
\begin{equation}
|Q|^2 = \sum_i |Q_i|^2 \ .
\end{equation}
In this case, the full effective Lagrangian is
\begin{eqnarray}
&& K \ = \ - \ 3 \ \ln \left(1 \ + \ X \ + \ {\bar X} \ - \ |Q|^2\right) \ , \nonumber \\
&& W \ = \ W_0 (Q_i) \ + \ f (Q_i) \, X  \ , \label{noscale2}
\end{eqnarray}
to be supplemented with the nilpotency condition $X^2=0$.
The corresponding scalar potential is
\begin{equation}
V \ = \ \frac{1}{3 (1-|Q|^2)^2} \left\{ \sum_i \left|\frac{\partial W_0}{\partial Q_i} \, +\,
{\bar Q_i} \, f \right|^2  \ + \ (1\ -\ |Q|^2) \, |f|^2 \ - \ 3 ({\bar f} \, W_0 \, +\, f \, {\bar W_0} ) \right\}
\ , \label{noscale3}
\end{equation}
and now the condition
\begin{equation}
a^{-1} \ = \ 3 \ \left(1 \ - \ |Q|^2\right) \ > \ 0 \quad \longrightarrow \quad |Q|^2 \ < \ 1 \  \label{noscale4}
\end{equation}
must clearly hold, in agreement with the preceding arguments. In general, these scalar potential are not positive definite, but positive definite potentials obtain if the relation
\begin{equation}
\sum_i Q_i \, \frac{\partial W_0}{\partial Q_i} \ - \ 3\, W_0 \ = \ - \ \frac{1}{2} \ f (Q_i) \label{noscale5}
\end{equation}
holds between the two functions entering the superpotential. Restricting the attention, for simplicity, to cubic superpotentials and imposing the condition
that Supersymmetry breaking occur along the sgoldstino direction, one finds the solution
\begin{eqnarray}
W_0 &=& \alpha \ + \ \frac{1}{2}\ b_{ij} \, Q_i \, Q_j \ + \ \frac{1}{6} \ \lambda_{ijk} \ Q_i \, Q_j \, Q_k \ ,  \nonumber \\
f &=& 6\, \alpha \ + \  b_{ij}\,  Q_i \, Q_j \ . \label{noscale6}
\end{eqnarray}
The scalar potential becomes in this case
\begin{equation}
V \ = \ \frac{1}{3 (1 \ - \ |Q|^2)^2} \ \sum_i \left|\frac{\partial W_0}{\partial Q_i} \right|^2
\ . \label{noscale7}
\end{equation}
For the vacuum there is always the solution $\langle Q_i \rangle = 0$, in which case
$D_i W = 0$ and $D_X W = 3 \alpha$. For vanishing \emph{v.e.v.}'s of the matter fields, the positivity of the potential can be ascribed to a cancelation between the terms $K^{X \bar X} |D_X W|^2$ and $- 3|W_0|^2$. However, the ansatz
(\ref{noscale6}) guarantees the positivity for arbitrary \emph{v.e.v.}'s of the matter fields that satisfy eq.~(\ref{noscale4}).

\subsection{\sc  \bf Some inflationary models}\label{sec:inflation}

Many inflationary models in Supergravity possess a shift symmetry in the inflaton multiplet $\Phi$. Simple models of inflation coupled to matter that possess a gravity dual can be constructed starting from the K\"ahler potential
\begin{eqnarray}
&& K_0 \ = \ - \ 3 \ \ln \left(1 \ - \ \frac{1}{2} \left(\Phi+{\bar \Phi}\right)^2 \ - \ |Q|^2 \right) \equiv
- \ 3\, \ln \ {\cal Z} \ , \nonumber \\
&& W \ = \ W_0 (\Phi,Q_i) \ + \ f (\Phi, Q_i) \, X  \
 \label{inf1}
\end{eqnarray}
where, as in Section \ref{sec:potentials},
\begin{equation}
|Q|^2 \ = \ \sum_i \, |Q_i|^2 \ .
\end{equation}

In this case, the full K\"ahler potential is given by
\begin{equation}
K \ = \ - \ 3 \ \ln \left(1 \ + \ X \ + \ {\bar X} \ - \ \frac{1}{2} \ (\Phi+{\bar \Phi})^2 \ - \
|Q|^2\right) \  , \label{inf2}
\end{equation}
to be supplemented with the nilpotency condition $X^2=0$. The inflaton $\varphi$ is identified with the (canonically normalized) imaginary
part of the complex field
\begin{equation}
\Phi = \frac{1}{\sqrt{6}}\ \left(\chi \ + \ i\ \varphi\right) \ ,
\end{equation}
where $\chi$ has a Hubble scale mass and is stabilized to zero during inflation, whereas $Q_i$ are matter fields.
The scalar potential has the general form (\ref{m12}), where the parameter $a$ is now given by
\begin{equation}
a^{-1} \ = \ \frac{3 {\cal Z}}{1 \ + \ \frac{1}{2} \left(\Phi+{\bar \Phi}\right)^2} \ > \ 0 \quad \longrightarrow
\quad \frac{1}{2} \, \left(\Phi+{\bar \Phi}\right)^2 \ + \ \sum_i |Q_i|^2 \ < \ 1 \ , \label{inf3}
\end{equation}
where ${\cal Z}$ was defined in eq.~\eqref{inf1}. The scalar potential is in this case
\begin{eqnarray}
V &=& \frac{1}{3 Y^2} \left\{ \sum_i \left|\frac{\partial W_0}{\partial Q_i} \, +\,
{\bar Q_i} \, f \right|^2   \ + \ \left(1\ +\ \frac{1}{2}\, \left(\Phi+{\bar \Phi}\right)^2 -\ |Q|^2\right) |f|^2 \
+ \ \left|\frac{\partial W_0}{\partial \Phi}\right|^2 \right. \nonumber \\
&+&\left. (\Phi+{\bar \Phi}) \left({\bar f} \, \frac{\partial W_0}{\partial \Phi}
 \ + \ {\rm h.c}\right)
 - \ 3\, ({\bar f} \, W_0 \ +\ f \, {\bar W_0} ) \right\}
\ . \label{inf4}
\end{eqnarray}
Inflationary models with a positive energy density during inflation can be found starting from the condition (\ref{noscale5}).
The solution is then
\begin{eqnarray}
W_0 &=& \alpha (\Phi) \ + \ \frac{1}{2}\ b_{ij} \, Q_i \, Q_j \ + \ \frac{1}{6} \ \lambda_{ijk} \ Q_i \, Q_j \, Q_k \ ,  \nonumber \\
f &=& 6\, \alpha (\Phi) \ + \  b_{ij}\,  Q_i \, Q_j \ , \label{inf5}
\end{eqnarray}
where $\alpha$ is now an appropriate function of the inflaton field.
The scalar potential in this case becomes
\begin{equation}
V \ = \ \frac{1}{3 Y^2} \ \left\{ \sum_i \left|\frac{\partial W_0}{\partial Q_i} \right|^2 \,+\,
\left|\frac{\partial W_0}{\partial \Phi}\right|^2 \, + \,  \frac{1}{2}\ (\Phi+{\bar \Phi})^2  \, |f|^2
\,+\,  (\Phi+{\bar \Phi}) \left({\bar f} \,\frac{\partial W_0}{\partial \Phi}\ + \ {\rm h.c}\right)  \right\}
\ . \label{inf6}
\end{equation}
Notice that matter fields and the inflaton have positive definite contribution to the scalar potential, while the non-positive definite part is proportional to the field $\chi$, which has to be massive and with vanishing \emph{v.e.v.} during inflation.
For vanishing \emph{v.e.v.}'s of the matter fields, the positivity of the potential can be ascribed, as in the previous section, to the no--scale structure \cite{noscale,noscale_2}. Indeed it results from a cancelation between the terms $K^{X \bar X} |D_X W|^2$ and $- 3|W_0|^2$, while the inflaton potential comes from $|D_{\Phi} W|^2$. The ansatz
(\ref{inf5}) ensures however positivity for arbitrary \emph{v.e.v.}'s of the inflaton and matter fields, satisfying (\ref{inf3}) with $\chi=0$.

This class of models is thus a natural generalization of corresponding ones in Section \ref{sec:potentials}. If the vacuum energy vanishes after inflation, matter fields with zero \emph{v.e.v.}'s and superpotential (\ref{inf5}) give a vanishing contribution to the energy at the minimum.
Appropriate choices for the functions $\alpha (\Phi)$ can then implement various inflationary models. For example, the choice
\begin{equation}
\alpha (\Phi) \ = \ \lambda \ - \ \frac{M}{2} \ \Phi^2 \  \label{inf7}
\end{equation}
leads to chaotic inflation with an inflaton potential $V (\varphi) = \frac{M^2 \varphi^2}{18}$.
The $\chi$  mass  during inflation $(\varphi>>1)$  is $m_\chi^2\approx \frac{M^2\varphi^4}{6}$, which is larger than the Hubble parameter $H$, so that $\chi$ can indeed be stabilized to zero during and after inflation. This model is similar to the one constructed in \cite{sugra_inflation_2}.

Cosmological inflation with a tiny tensor--to--scalar ratio $r$, consistently with PLANCK data, may also be described within the present framework, for instance choosing
\begin{eqnarray}
 \alpha(\Phi)\ =\ i\, M\left( \Phi\ +\ b \, \Phi \, e^{\,ik\, \Phi}\right) \ .
 \end{eqnarray}
This potential bears some similarities with the K\"ahler moduli inflation of \cite{kahler_mod_inf} and with the poly--instanton inflation of \cite{poly_mod_inf}.
One can verify that $\chi=0$ solves the field equations, and that the potential along the $\chi=0$ trajectory is now
\begin{eqnarray}
 V\ =\ \frac{M^2}{3}\left(1 \ -\  a \ \phi \ e^{-\,\gamma \,\phi}\right)^2\ .
 \end{eqnarray}
Notice that we have defined, for brevity,
\begin{equation}
\phi \ = \ \varphi \ - \ \frac{\sqrt{6}}{k} \ ,
\end{equation}
while
\begin{equation}
a\ =\ \frac{b\,\gamma}{e} \ < \ 0 \ , \qquad \gamma\ =\ \frac{k}{\sqrt{6}} \ < \ 0 \ .
\end{equation}
The potential is clearly positive definite and has a minimum determined by the condition
\begin{equation}
a\, \phi\ =\ e^{\,\gamma\,\phi} \ ,
\end{equation}
which can be solved for $\phi$ in terms of a Lambert $W$ function \cite{lambert}. This is a global minimum, and corresponds to a non--supersymmetric Minkowski vacuum satisfying  eq.~(\ref{m13}), \emph{i.e.},
\begin{eqnarray}
\langle \, f\, \rangle\ = \ \left(3\ + \ \sqrt{\frac{3}{a}}\right)\ \langle \, e^{\,\frac{1}{3}\,K_0}\ W_0\, \rangle \ .
\end{eqnarray}

In this model inflation occurs for large negative values $\phi<< \frac{1}{\gamma}$ where the potential develops a plateau and takes approximately the form
\begin{eqnarray}
V\ \approx \ \frac{M^2}{3}\left(1\ -\ 2\,a \, \phi \ e^{-\gamma \phi}\right)\ .
\end{eqnarray}
The key inflationary parameters, the spectral index $n_s-1$ and the tensor--to--scalar ration $r$, are in this case
\begin{eqnarray}
n_s-1\ \simeq \ -\ \frac{2}{N}\ , \qquad r\ \simeq \ \frac{8}{N^2 \, \gamma^2}\ , \label{par}
\end{eqnarray}
and are thus consistent with the latest PLANCK data \cite{PLANCK} for $|\gamma| > \frac{9}{N}$.
Note also that the mass of the  $\chi$ field  is always positive, for any value of $\phi$, so that $\chi=0$ is indeed a stable trajectory.
In fact, for large negative  values of $\phi$
\begin{eqnarray}
   m_\chi^2\ \simeq \ \frac{4(3\ +\ \gamma^2)}{3\, \gamma^2} \ a^2\ M^2 \ \phi^2\ e^{\,-2\,\gamma\, \phi} \qquad \left(\phi\ <<\ \frac{1}{\gamma}\right) \ .
\end{eqnarray}
Therefore, $m_\chi\ >>\ H\ =\ \frac{M}{3}$ during inflation, so that the $\chi$ field  indeed decouples, and the inflationary dynamics is well described by the single
field $\phi$ as it undergoes slow roll along the potential from large negative values, with the parameters given in eq.~(\ref{par}).


\vskip.3in
\noindent
{\bf Acknowledgments}

\noindent We are grateful to I.~Antoniadis, R.~Kallosh, M.~Porrati, A.~Van Proeyen and F.~Zwirner for discussions. The work of E.~D. was partly supported by a Humboldt Research Award 2014 at DESY. S.~F is supported in part by INFN (I.S. GSS). A.~S. is on sabbatical leave, supported in part by Scuola Normale and by INFN (I.S. Stefi), and would like to thank the CERN Th-Ph Department for the kind hospitality.

\end{document}